\documentclass[conference]{IEEEtran}
\IEEEoverridecommandlockouts
\usepackage{cite}
\usepackage{amsmath,amssymb,amsfonts}
\usepackage{algorithmic}
\usepackage{graphicx}
\usepackage{textcomp}
\usepackage{comment}
\usepackage{xcolor}
\usepackage[inline]{enumitem}
\def\BibTeX{{\rm B\kern-.05em{\sc i\kern-.025em b}\kern-.08em
    T\kern-.1667em\lower.7ex\hbox{E}\kern-.125emX}}
\begin{document}

\title{Layer 2 Coordinated Trusted Setup for Continuous CRS Generation\\
}
\author{
Khalid Hassan$^{1}$,
Sara Rouhani$^{1,2}$\\
$^{1}$Department of Computer Science, University of Manitoba, Winnipeg, MB, Canada\\
$^{2}$Department of Software and IT Engineering, École de technologie supérieure (ÉTS), Montréal, QC, Canada\\
\texttt{hassank2@myumanitoba.ca, sara.rouhani@umanitoba.ca, sara.rouhani@etsmtl.ca}
}

\maketitle

\begin{abstract}
Zero-knowledge proof systems rely on a trusted setup phase to generate a Common Reference String (CRS), yet existing approaches are typically static, one-time ceremonies that are inflexible and vulnerable to long-term compromise. Offloading continuous, recurring trusted setups to a decentralized Layer 2 (L2) network introduces a fundamental coordination challenge arising from the mismatch between high-throughput transaction processing and the multi-round requirements of trusted setup ceremonies. This paper presents an L2-coordinated framework that safely decouples transaction pipelines from ceremony execution to achieve automated, continuous CRS generation without centralized coordination. We design and implement two protocol variants over a decentralized, PBFT-coordinated ZK-rollup architecture: an on-chain smart contract approach and an asynchronous peer-to-peer consensus variant. Both designs utilize non-interactive zero-knowledge proofs of knowledge alongside commit-reveal structures to eliminate adaptive manipulation vectors and isolate ceremony latency. Experimental evaluations under simulated wide-area network constraints and adversarial conditions demonstrate that our architecture successfully isolates ceremony liveness. Continuous setups complete reliably within practical time bounds despite node dropouts or malicious contributions, while preserving stable L2 transaction throughput.

\end{abstract}


\section{Introduction}

Blockchain networks have long grappled with the structural trade-offs between scalability, security, and decentralization—commonly referred to as the Blockchain Trilemma \cite{Werth2023ARO}. While decentralization eliminates single points of failure, throughput constraints on Layer 1 (L1) networks remain a persistent challenge, with Ethereum processing roughly 15 transactions per second (TPS) and Bitcoin averaging around 7 TPS, compared to conventional financial infrastructures such as Visa \cite{Buterin2013,TRIPATHI2023100344}. To address these limitations, Layer 2 (L2) rollups have emerged as a promising scaling solution by executing transactions off-chain, batching them, and anchoring state updates to L1 for security and finality \cite{gangwal2023survey,pareek2025layer}. Beyond scalability, recent studies have highlighted the growing role of L2 infrastructures in supporting broader blockchain ecosystem functionality, including interoperability and cross-network coordination \cite{hassan2025sok}.

Rollups generally fall into two categories: optimistic rollups, which rely on fraud proofs, and zero-knowledge (ZK) rollups, which use cryptographic proofs to validate state transitions. However, despite the broader decentralization goals of blockchain systems, most existing rollup architectures continue to rely on centralized sequencers, reintroducing a single point of control and failure \cite{motepalli2023sok}.

Concurrently, zero-knowledge (ZK) systems rely on a Common Reference String (CRS) for proof generation and verification. To minimize trust assumptions, the CRS is typically generated collaboratively through multi-party computation (MPC) protocols such as Powers of Tau (PoT) \cite{tau}. Although these trusted setup ceremonies are designed to produce a secure CRS as long as at least one participant behaves honestly, practical deployments often require the CRS to be regenerated over time. Such regeneration may be necessary when circuits evolve, participants join or leave the system, governance policies change, or long-term trust assumptions need to be refreshed. However, existing trusted setup protocols are primarily designed as one-time ceremonies with fixed participant sets, making repeated ceremonies costly and difficult to adapt to these changing requirements.

This creates an open challenge: how can trusted setup be transformed from an isolated cryptographic event into a continuously available decentralized service while preserving the security guarantees of multi-party CRS generation? While decentralized sequencer architectures have been proposed to improve transaction ordering robustness and reduce reliance on centralized operators \cite{capretto2025decentralized,chonky,bearer2024espresso}, they have not been explored as a coordination substrate for recurring CRS lifecycle management. Furthermore, rollups and trusted setup ceremonies exhibit fundamentally different operational characteristics. Rollups are optimized for high-throughput transaction processing and low-latency state transitions, whereas trusted setup ceremonies involve multi-round coordination, participant synchronization, and computationally intensive cryptographic operations. A naive integration of the two can degrade system performance and introduce opportunities for strategic manipulation during ceremony execution.

To address these challenges, this paper proposes a L2 coordinated framework for continuous CRS generation. Rather than treating trusted setup as an isolated one-time event, we leverage a decentralized sequencer network to coordinate recurring CRS ceremonies and support decentralized CRS lifecycle management. The framework supports both smart contract-based and peer-to-peer coordination models and enables periodic CRS regeneration using rotating subsets of sequencers. This approach allows CRS generation to operate as a recurring protocol service while remaining decoupled from the transaction processing pipeline. We evaluate the proposed framework along three dimensions: transaction throughput of the underlying L2 network, completion time and robustness of CRS generation ceremonies under adversarial conditions, and the size of the generated CRS.
The primary contributions of this paper are as follows:
\begin{enumerate}

\item \textbf{Layer 2 Coordination Framework for Continuous Trusted Setup.}
We design a decentralized coordination framework that leverages a sequencer network to transform trusted setup from a one-time ceremony into a recurring protocol service. The framework enables decentralized CRS lifecycle management through repeated generation, verification, publication, and replacement of CRS instances.

\item \textbf{Continuous CRS Generation Protocol.}
We propose a decentralized CRS generation protocol integrated within a rollup-based L2 environment. The protocol supports both smart contract-based and peer-to-peer coordination models and enables periodic CRS regeneration using rotating subsets of sequencers. By decoupling ceremony execution from transaction processing, the design supports recurring CRS generation while preserving system liveness and reducing long-term reliance on a fixed participant set.

\item \textbf{Implementation and Empirical Evaluation.}
We implement the proposed framework on a rollup-based Layer 2 platform and evaluate it under varying participant sizes and adversarial conditions. Results show that recurring CRS ceremonies complete reliably within practical time bounds while maintaining stable throughput.

\end{enumerate}

\section{Background and Related Work}\label{background}

\subsection{Sequencer Centralization in Rollups}

Rollups \cite{thibaultRollups22} have emerged as the dominant L2 scaling solution for Ethereum, improving throughput and reducing transaction costs. They are generally classified into Optimistic Rollups \cite{armstrong2021ethereum} and zero-Knowledge Rollups (ZK-Rollups) \cite{chaliasos2024analyzing}. Optimistic rollups assume transactions are valid unless challenged through fraud proofs during a dispute period, whereas ZK-rollups generate cryptographic proofs that allow transaction validity to be verified directly on L1. While the former reduces computational overhead, the latter eliminates challenge periods at the cost of proof generation.

Regardless of their type, most rollups rely on centralized sequencers for transaction ordering and inclusion. This creates a single point of failure and introduces censorship, delay, and soft-finality risks \cite{koegl2023attacks, motepalli2023sok}. To address these concerns, several decentralized sequencing architectures have been proposed. Capretto et al. introduce \textit{arrangers}, which combine sequencing and data availability through decentralized consensus \cite{capretto2025decentralized}. Similarly, zkSync has proposed ChonkyBFT \cite{chonky}, a scalable BFT protocol that reduces communication overhead through threshold signatures, while Espresso \cite{bearer2024espresso} provides a production-grade shared sequencer enabling neutral ordering across multiple rollups.

While these approaches improve transaction ordering robustness, they primarily target sequencing and data availability. In contrast, our work leverages a decentralized sequencer network as a coordination layer for recurring cryptographic ceremonies, specifically continuous CRS generation. The proposed framework is compatible with existing decentralized sequencing architectures and extends utility beyond transaction processing to support fault-tolerant trusted setup services.
\subsection{ZKP and Multi-party Setups}

Zero-knowledge proofs (ZKPs) are a key cryptographic primitive for enhancing privacy in blockchain systems \cite{promise}. While blockchains provide transparency and auditability, they can expose sensitive transaction information, motivating the use of privacy-preserving mechanisms \cite{bernabe2019privacy}. Among the various ZKP constructions, zero-Knowledge Succinct Non-Interactive Arguments of Knowledge (zkSNARKs) are widely adopted due to their compact proof sizes and efficient verification \cite{zkpsurvey}.

Most zkSNARK systems require a trusted setup phase to generate a Common Reference String (CRS), which contains the public parameters used for proof generation and verification. Because the CRS underpins the security of the system, a compromised setup may enable the creation of fraudulent proofs \cite{scalablezkp}.

To mitigate this risk, Powers of Tau (PoT) \cite{tau} employs a multi-party computation (MPC) ceremony in which multiple participants contribute randomness to the setup process. Security is preserved as long as at least one participant behaves honestly and securely discards its secret contribution. PoT first generates a universal CRS that can be reused across a broad class of zkSNARK circuits, after which circuit-specific parameters can be derived while maintaining the security guarantees of the setup.
\section{System Design}\label{design}
\subsection{Architectural Overview}

The proposed architecture consists of two core components. The first is a decentralized sequencer network coordinated using PBFT consensus, which provides fault-tolerant transaction ordering and eliminates the single point of failure associated with centralized rollup sequencers.

Built on this coordination layer, the second component is a decentralized CRS generation service based on the PoT protocol. In each generation round, a randomly selected subset of sequencers participates in a multi-party computation process to produce a fresh CRS instance. The framework supports both smart contract–based and peer-to-peer coordination models. In either case, the resulting CRS is anchored to the L1 blockchain to provide public verifiability and integrity.

\subsection{Decentralized Sequencer Model}

\begin{figure}[htbp]
    \centering
    \includegraphics[height=150px]{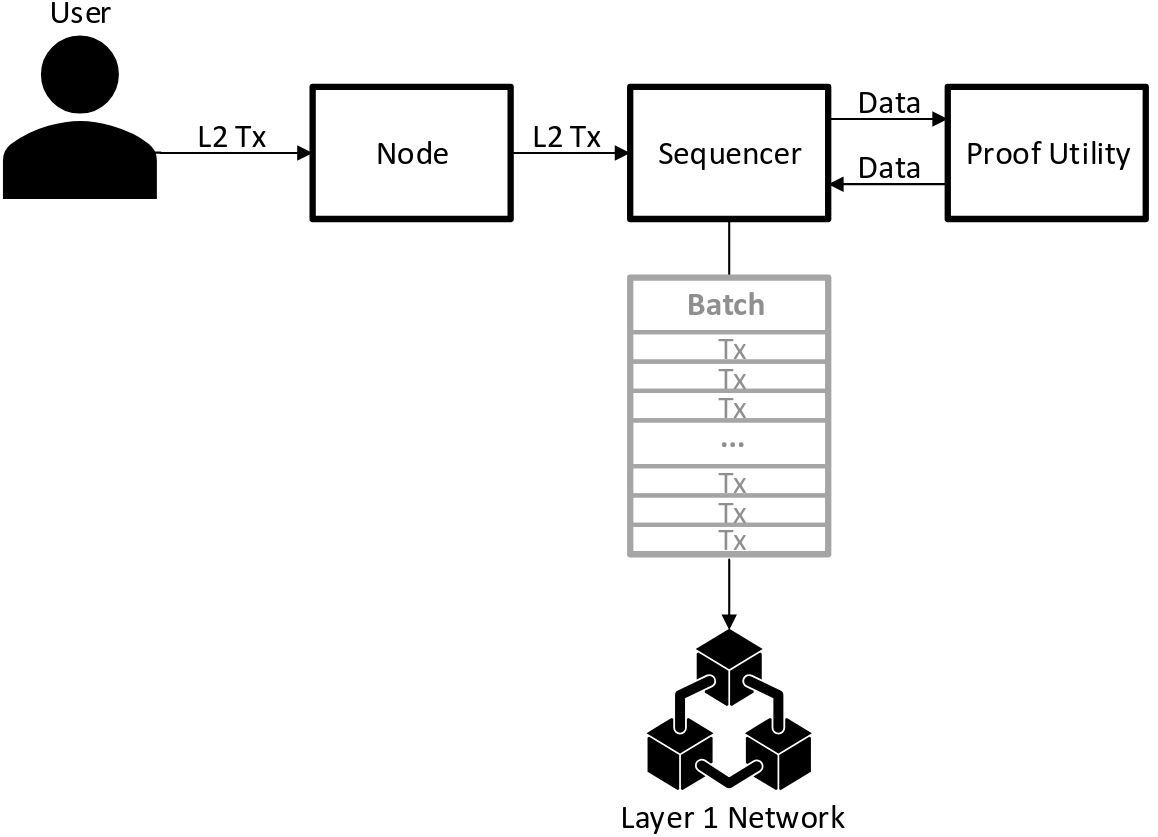}
    \caption{Centralized Sequencer Architecture}
    \label{fig:centralized_sequencer}
\end{figure}

\begin{figure}[htbp]
    \centering
    \includegraphics[height=200px]{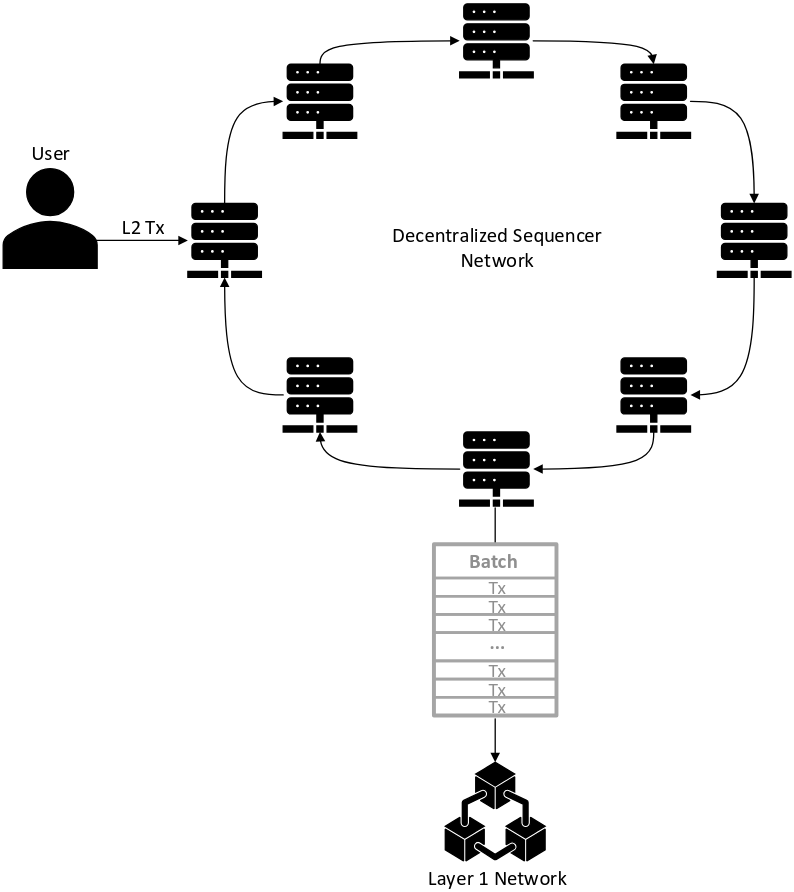}
    \caption{Decentralized Sequencer Architecture}
    \label{fig:decentralized_sequencer}
\end{figure}

The sequencer is a core component of a rollup system, responsible for transaction ordering and batch formation. In conventional architectures, transactions are relayed to a centralized sequencer as presented in Figure \ref{fig:centralized_sequencer}, creating a single point of failure and control. In contrast, a decentralized sequencer model replaces the centralized entity with a peer-to-peer (P2P) network of nodes that collectively agree on transaction ordering through a consensus protocol, as illustrated in Figure \ref{fig:decentralized_sequencer}.

Beyond improving fault tolerance, decentralized sequencing distributes control over transaction inclusion and ordering across multiple participants, reducing censorship risks and increasing resilience against node failures and targeted attacks. Consensus-based coordination also provides a globally consistent view of system state, enabling sequencers to agree on protocol outcomes despite network delays or Byzantine behavior within the supported fault threshold. These properties make decentralized sequencer networks suitable not only for transaction ordering but also for coordinating distributed services that require shared state and ordered execution.

The rollup architecture developed in this work leverages these properties by using the sequencer network as both the transaction ordering layer and the coordination substrate for recurring CRS generation. Sequencer nodes communicate through a P2P networking layer that supports transaction propagation, batch dissemination, and consensus coordination. Agreement is achieved using Practical Byzantine Fault Tolerance (PBFT), which proceeds through the standard \textit{Pre-prepare}, \textit{Prepare}, and \textit{Commit} phases. A batch is finalized once a supermajority of nodes approves it.

To encourage honest participation, the framework incorporates a reputation mechanism that penalizes disruptive behavior. Minor faults, such as occasional disconnections, incur small reputation penalties, while repeated failures or malicious actions, including invalid CRS contributions, result in larger penalties and eventual exclusion from future consensus and CRS generation rounds. During an active ceremony, participants that timeout or submit invalid contributions are immediately removed from the current participant set to preserve ceremony liveness.

Once a batch is finalized, a zkSNARK proof is generated to attest to the correctness of the resulting state transition. The proof, state root, and associated batch metadata are then submitted to an L1 smart contract, which verifies the proof and anchors the batch on-chain.

\subsection{Decentralized CRS Generation}
ZKPs are cryptographic protocols that enable one party, the prover, to convince another party, the verifier, of the validity of a statement without revealing any additional information. Among the various classes of ZKPs, zkSNARKs (Zero-Knowledge Succinct Non-Interactive Arguments of Knowledge) are particularly well-suited for blockchain applications due to their compact proof sizes and efficient verification. These properties make zkSNARKs ideal for enabling privacy-preserving transactions, scalable on-chain computation, and secure multi-party protocols.

Many zkSNARK systems rely on a Common Reference String
(CRS), which serves as a publicly shared parameter set
required for proof generation and verification. For
illustrative purposes, CRS transformations are represented
abstractly as multiplicative updates over elliptic curve group
elements. Under this abstraction, the CRS can be represented
as a tuple $(G,H)$, where $G$ and $H$ are sets of structured
points on elliptic curves. These points encode the public
parameters used during proof generation and verification.
Specifically:

\[
G = \{G_1, G_2, ..., G_n\}, \quad
H = \{H_1, H_2, ..., H_m\}
\]

where $G_i$ and $H_j$ are elements of an elliptic curve group,
and the values of $n$ and $m$ depend on the size and complexity
of the underlying zkSNARK construction.

The security of zkSNARKs is fundamentally tied to the integrity of the CRS. If the CRS is generated maliciously or manipulated, an adversary may be able to forge proofs that appear valid to verifiers. To mitigate this risk, the CRS must be generated in a manner that prevents any single participant from controlling the output. This is typically achieved through multi-party computation (MPC) protocols, in which multiple independent participants contribute randomness to collaboratively construct a secure CRS.

In a multi-party CRS generation ceremony, each participant applies a transformation to the CRS using a secret random scalar \( r_i \). Mathematically, the transformation is expressed as: 

\[ G^{'}_i = r_i \cdot G_i, H^{'}_j = r_i \cdot H_j \]

where \( G^{'}_i \) and \( H^{'}_j \) represent transformed points, and \( r_i \) is a scalar randomly sampled from the underlying elliptic curve group. The final CRS is computed by sequentially applying each participant’s transformation in turn:

\[ G_{final} = r_i \cdot (r_{i-1} \cdot (...(r_1 \cdot G)...)), H_{final} = r_i \cdot (r_{i-1} \cdot (...(r_1 \cdot H)...)) \]

This process ensures the unpredictability of the final CRS. A key security property is that the CRS remains cryptographically secure as long as at least one participant behaves honestly and keeps their scalar \( r_i \)  secret. The security of the construction relies on the hardness of the elliptic curve discrete logarithm problem (ECDLP), which makes it computationally infeasible to determine a scalar \( r \) such that \( B = r \cdot A\) where \( A \) and \( B \) are both points on the curve, ensuring that the individual contributions of participants cannot be reversed, even if all other participants collude.

While CRS generation has been deployed in real-world systems such as PoT, existing approaches are typically coordinated as standalone ceremonies and can incur significant operational overhead. To address this limitation, we leverage a decentralized Layer 2 sequencer network as a coordination substrate for recurring CRS generation while preserving the trust-minimized and distributed nature of the ceremony. The framework supports two coordination models: \begin{enumerate*} \item smart contract–based coordination and \item peer-to-peer (P2P) coordination among sequencer nodes. \end{enumerate*}
\subsubsection{On-chain Coordinated CRS Generation}
Using the previously detailed multi-party CRS generation scheme, this design uses a smart contract deployed on L2 to manage the entire ceremony state. Furthermore, the system employs a commit-reveal scheme during the contribution process. Each participant first commits to their random scalar \( r_i \) by publishing a hash \( h_i = H(r_i) \), where \( H \) is a cryptographic hash function. Crucially, the secret scalar $r_i$ is never exposed to the smart contract, the sequencers, or any other participant; it is immediately and permanently deleted by the local node upon generating the contribution to eliminate the \textit{toxic waste} trapdoor. This framework prevents participants from front-running or colluding to shape the final parameters, thereby preserving the cryptographic security and integrity of the CRS.

CRS generation rounds occur periodically at a predetermined interval agreed upon by the participating nodes. In each round, selected participants register, submit commitments, and sequentially contribute to the transformation of the CRS. Once all contributions have been applied, the resulting CRS is finalized, stored, and made available for subsequent zkSNARK proof generation and verification. The full procedure can be summarized as follows:

\begin{enumerate}
    \item \textit{Registration}: Nodes register themselves in a pool of eligible participants, from which a subset  \( P = \{P_1,P_2,...,P_k\} \) is selected in each round to participate in the CRS generation ceremony.
    \item \textit{Commitment}: Each participant \( P_i \) from the set of participants \( P \) generates a random scalar \( r_i \), computes a cryptographic commitment \( h = H(r_i) \), and submits to the system to prevent premature disclosure.
    \item \textit{Contribution}: Once all commitments have been submitted, participants sequentially apply their contributions. Each participant \( P_i \) updates the CRS as follows: \( G_i=r_i \cdot G_{i-1}, H_i=r_i\cdot H_{i-1} \).
    \item \textit{Finalization}: Once all contributions have been applied, the final CRS \( (G_{final}, H_{final}) \) is stored and commitments are revealed to ensure the integrity of the final generated CRS.
\end{enumerate}

\subsubsection{Fully-decentralized CRS Generation}

In this design, CRS generation is executed directly by the sequencer network without relying on smart contracts. Coordination is achieved through a shared ceremony state replicated across participating nodes, which tracks the current round, participant set, generation step, and intermediate CRS.

The ceremony is integrated with the PBFT protocol to maintain a consistent CRS state across the network. During each generation round, randomly selected participants sequentially apply transformations to the intermediate CRS. Following each contribution, the updated state is validated and agreed upon through PBFT before the next participant proceeds. Once all contributions have been applied, the final CRS is generated and anchored to the L1 chain.

Each generated CRS is assigned a unique epoch identifier and anchored to the Layer~1 blockchain to provide an immutable record of its validity. Transactions and proofs generated during epoch $e$ use the corresponding CRS, denoted by $\mathrm{CRS}_e$. A newly generated CRS, $\mathrm{CRS}_{e+1}$, becomes active only after successful completion of the ceremony and confirmation of its Layer~1 anchor. Previous CRS versions remain available for proof verification until all pending proofs referencing $\mathrm{CRS}_e$ have been finalized, after which they are retired. This epoch-based lifecycle management ensures seamless CRS transitions while preserving compatibility with in-flight proof generation and verification.


\section{Security Analysis} \label{SecurityAnalysis}

This section defines the threat model, security assumptions, and key security properties of the proposed framework. We show how the protocol preserves the trust guarantees of PoT-style ceremonies while improving operational resilience through decentralized coordination, recurring CRS generation, and Layer 1 anchoring.

\subsection{Threat Model}

We consider a probabilistic polynomial-time adversary that may control a subset of sequencer nodes and behave maliciously during CRS generation. Adversarial actions include submitting invalid contributions, deviating from the protocol, disconnecting during execution, colluding with other participants, and attempting to bias the ceremony outcome through adaptive behavior.

We assume the adversary cannot violate standard cryptographic assumptions, including the hardness of the elliptic curve discrete logarithm problem (ECDLP) and the collision resistance of the commitment hash function. At the network level, we consider temporary message delays and partial communication failures. These are mitigated by the underlying PBFT protocol, which provides ordered execution and fault-tolerant agreement within its standard fault threshold.

 \subsection{Security Assumptions}

The proposed scheme follows the standard trust model of PoT-style multi-party trusted setup ceremonies \cite{tau}. In each ceremony round, participating sequencers sequentially apply secret scalar transformations to an intermediate CRS. Security relies on the assumption that at least one participant in a given round behaves honestly and does not reveal or retain its secret randomness after contributing.

Under this assumption, the final CRS remains secure even if all other participants are malicious or colluding. More specifically, as long as one contribution is honestly generated and its secret randomness is destroyed, no adversary can reconstruct the effective trapdoor embedded in the CRS. This assumption is standard in sequential contribution ceremonies and is inherited by our protocol.

\subsection{CRS Security Properties}

During each ceremony round, sequencers sequentially update the intermediate CRS through multiplicative secret-scalar contributions. The resulting CRS therefore depends on the composition of all participant contributions rather than on any single sequencer. Under the ECDLP assumption, recovering the effective trapdoor from the public CRS remains computationally infeasible unless the standard trusted setup assumption is violated.

This property ensures that disclosure, corruption, or later compromise of a subset of participants does not by itself invalidate the generated CRS. The security of the final CRS is broken only in the extreme case where all contributions in a ceremony round are exposed or controlled by colluding adversaries. Thus, the protocol preserves the standard PoT security guarantee while embedding it into a decentralized coordination framework.

\subsection{Commit--Reveal Protocol}

The commit--reveal mechanism strengthens the ceremony against adaptive adversarial behavior. In the commitment phase, each participant first publishes a cryptographic commitment to its contribution without revealing the underlying randomness. Only after all commitments have been fixed are the corresponding values revealed and verified.

This mechanism prevents an adversary from choosing its contribution after observing the randomness of honest participants, thereby mitigating selective bias and last-mover influence over the final CRS. In particular, it binds each participant to a predetermined contribution while preserving secrecy during the commitment stage. As a result, the final CRS remains less susceptible to manipulation through reactive contribution strategies.

\subsection{Resilience to Compromise and Collusion}

\emph{temporally bounded compromise}: Let CRS$_k$ denote the CRS generated in ceremony round $k$. We say that the protocol achieves \emph{temporally bounded compromise} if the security of CRS$_k$ depends only on the participant set and execution of round $k$, and not on prior or future rounds. In particular, even if all participants in round $k$ are adversarial and the corresponding CRS$_k$ is compromised, the security of CRS$_{k+1}$ is preserved as long as the standard trusted setup assumption holds in round $k+1$, i.e., at least one participant behaves honestly and its secret randomness remains unknown.

This property follows from the independence of ceremony rounds and the use of fresh randomness and participant subsets in each round. Consequently, compromise of one ceremony is confined to the corresponding CRS instance and does not affect future generations. While CRS regeneration does not retroactively repair compromised CRS instances, it limits future exposure by establishing fresh trust assumptions in subsequent rounds.

A central limitation of conventional trusted setup ceremonies is that they are executed as one-time events, so the trust assumption is tied permanently to a single participant set and a single execution instance. In contrast, the proposed design supports periodic CRS regeneration using rotating subsets of sequencer nodes. This transforms trusted setup from a static ceremony into a recurring protocol service.

As a result, the impact of a compromised ceremony round is temporally bounded. Even if all participants in one round were malicious or colluding, the affected CRS can later be replaced through a subsequent regeneration round involving a different participant subset. While this does not eliminate trust assumptions within any individual round, it reduces long-term reliance on a fixed set of participants and improves resilience to collusion over time.

\subsection{Verifiability and Anchoring}

The final CRS generated in each ceremony round is anchored to the Layer 1 blockchain, providing an immutable public record of the ceremony outcome. This anchoring improves transparency by allowing external observers to verify which CRS instance was finalized and committed for subsequent use.

In addition, contribution updates are subject to protocol-level validity checks before being incorporated into the final CRS state. Invalid or malformed contributions can therefore be detected during the ceremony and prevented from affecting the finalized output. Together, protocol verification at the coordination layer and immutable anchoring at Layer 1 provide a publicly auditable execution trail and reduce the risk of undetected tampering.

 
\section{Implementation}\label{implementation}

The implementations of the decentralized ZK-Rollup and the decentralized CRS generation\footnote{https://github.com/wiiatcrs/wiiatzk} can be found on GitHub.

\subsection{Sequencer Decentralization}
The implementation of the proposed decentralized ZK-Rollup is developed in Golang and adheres to the architectural design outlined in the previous section. The system is modularized into distinct components, each encapsulated in a dedicated package to address a specific aspect of the rollup's functionality. These components include: \begin{enumerate*}
    \item P2P networking,
    \item Consensus,
    \item Rollup state management,
    \item Ethereum Virtual Machine (EVM) execution,
    \item and L1 integration
\end{enumerate*}.

The P2P  networking layer is implemented using the \textit{libp2p} library \cite{libp2p}, which provides well-established primitives for peer discovery, connection management, and message broadcasting. Each node is assigned a unique identity and binds to a designated port on the host system. For dynamic peer discovery, nodes rely on the Kademlia Distributed Hash Table (DHT) protocol \cite{dht}, ensuring decentralized connectivity and network resilience. The system registers three protocol handlers to process incoming messages: one each for transactions, batch propagation, and consensus coordination. These handlers facilitate the reliable exchange of information required for transaction ordering and batch construction across the network.

Building upon the P2P networking layer, the system implements a Practical Byzantine Fault Tolerance (PBFT) consensus mechanism to enable nodes to reach agreement on the ordering of transactions within each batch. The PBFT core manages consensus states, handles leader rotation, and processes all incoming consensus-related messages. A unified message structure, detailed in Table \ref{table:consensus_message}, is employed to ensure flexibility: the \textit{MessageType} field specifies the message role (e.g., \textit{PrePrepare}, \textit{Prepare}, or \textit{Commit}), while only the relevant fields are populated based on message type.

During consensus, the designated leader proposes a batch by broadcasting a \textit{PrePrepare} message containing the batch hash and associated metadata. Upon validation, the other nodes (followers) respond with \textit{Prepare} messages. Once a node collects a quorum of valid \textit{Prepare} messages, it issues a \textit{Commit} message to finalize the batch.

\begin{table}[htbp]
    \centering
    \caption{Structure of \texttt{ConsensusMessage}}
    \resizebox{\linewidth}{!}{%
    \begin{tabular}{| c | c |}
        \hline
        \textbf{Field Name} & \textbf{Description} \\
        \hline
        \texttt{Type} & Type of the message (Preprepare, Prepare, or Commit). \\
        \hline
        \texttt{View} & Current view number. \\
        \hline
        \texttt{Sequence} & Sequence number for the consensus round. \\
        \hline
        \texttt{BatchHash} & The hash of the proposed batch. \\
        \hline
        \texttt{NodeID} & The identifier of the node from which the message originates. \\
        \hline
        \texttt{Timestamp} & The timestamp at which the message was sent. \\
        \hline
        \texttt{Batch} & The proposed batch (only included in Preprepare). \\
        \hline
    \end{tabular}%
    }
    \label{table:consensus_message}
\end{table}
The sequencer integrates transaction propagation, consensus, and L1 anchoring into a unified workflow. Validated transactions are batched and finalized through consensus, after which ZK proofs (or dummy proofs) are generated and submitted to an Ethereum L1 contract for verification and anchoring.

State transitions resulting from finalized batches are maintained by the state management layer, which tracks accounts, balances, smart contract code, and storage. The state is represented through Merkle-tree-based state roots and exposed through a \textit{StateDB} interface that enables Ethereum-compatible execution.

Smart contract deployment and execution are supported through a simplified EVM execution layer, which interacts with the \textit{StateDB} to retrieve and update rollup state while maintaining compatibility with Ethereum semantics.

L1 integration is provided through an \textit{ethclient} component responsible for transaction signing, batch submission, and interaction with the rollup contract. Finalized batches, together with their corresponding state roots and proofs, are anchored on Ethereum for verification and persistence.

\subsection{Decentralized CRS Generation}

\subsubsection{CRS Generation using Smart Contracts}
The cornerstone of this scheme's implementation is the \textit{CRSManager} smart contract, paired with cryptographic utilities implemented in the sequencers for CRS contribution computation. Both the smart contract and the sequencers complement one another to manage the lifecycle of the CRS generation ceremony, ensure CRS integrity, and perform secure transformations to obtain a final CRS to be utilized by zkSNARK applications.

The smart contract serves as the coordination mechanism. for orchestrating the CRS generation process, maintaining the state of the current round, including the list of registered participants, deadlines, and the active CRS while enforcing access control, ensuring non-participants cannot meddle with the process. The contract enables participants to register for the ceremony and allows for commitment submission after a participant has been registered to prevent premature disclosure of contributions. Contributions are sequentially applied with each participant transforming the transient CRS using Circom's PoT implementation and submitting the result to the contract. Furthermore, the contract maintains the order of participants, ensuring that none other than the participant whose turn it is to contribute can do so. Once all contributions have been made, the last participant finalizes the round by calling the \textit{finalizeCRS} function, storing the final CRS on the ledger.

The CRS generation process begins with smart contract deployment, where parameters such as \textit{roundDuration} and \textit{maxParticipants} are defined. Participants then register, submit commitments, and sequentially contribute random transformations to the CRS. Once all contributions have been applied, the final CRS is stored and the system prepares for the next generation round.

\subsubsection{CRS Generation using P2P Networking}

The P2P-based CRS generation protocol is integrated directly into the PBFT consensus workflow through a shared \textit{CRSCeremonyState}, summarized in Table \ref{table:crs_cer_state}. Similar to the smart contract variant, CRS transformations are performed using Circom's PoT implementation.

During each ceremony round, participants sequentially contribute randomness through the \textit{AddContribution} function. Each proposed CRS update is independently verified against the previous ceremony state by the remaining participants. Upon supermajority approval, the shared ceremony state is updated with the new intermediate CRS and the protocol advances to the next contribution step. After all contributions have been applied, the final CRS is generated and anchored to the L1 chain.

\begin{table}[htbp]
    \centering
    \caption{Structure of \texttt{CRSCeremonyState}}
    \resizebox{\linewidth}{!}{%
    \begin{tabular}{| c | c |}
        \hline
        \textbf{Field Name} & \textbf{Description} \\
        \hline
        \texttt{EpochNumber} & The current epoch of the CRS generation ceremony. \\
        \hline
        \texttt{Participants} & The nodes participating in the current ceremony. \\
        \hline
        \texttt{CurrentStep} & The index indicating whose turn it is to contribute to the CRS. \\
        \hline
        \texttt{IntermediateCRS} & The incomplete CRS to which participants must contribute. \\
        \hline
        \texttt{Completed} & Indicates the completion status of the ceremony. \\
        \hline
    \end{tabular}%
    }
    \label{table:crs_cer_state}
\end{table}

\section{Evaluation}\label{evaluation}
The evaluation focuses on two aspects: (1) the transaction throughput of the decentralized sequencer network within the ZK-Rollup under stress testing and (2) the completion time and robustness of decentralized CRS generation ceremonies under network faults and malicious behavior.

All experiments were conducted using a simulated distributed environment hosted on a machine equipped with an Intel Xeon Silver 4216 processor (64 CPUs), 128 GB RAM, and 500 GB NVMe storage. Multiple sequencer nodes were deployed as independent processes communicating through the P2P networking layer, enabling evaluation of decentralized coordination and CRS generation behavior while controlling for hardware variability. 

\subsection{Decentralized ZK-Rollup}

The decentralized ZK-Rollup implementation is evaluated based on its transaction throughput. To perform the evaluation, three nodes are launched on the host machine to form a decentralized sequencer network and accept transactions. The system is then tested by submitting generated transaction batches with sizes varying from 20,000 transactions to 100,000 transactions. For each batch, real-world conditions are simulated by distributing the transactions across a fixed number of unique senders, ensuring that each sender’s nonce increments sequentially to avoid waiting for transaction finalization to obtain the latest nonce after consensus.

\begin{figure}[htbp]
    \centering
    \includegraphics[width=0.8\linewidth]{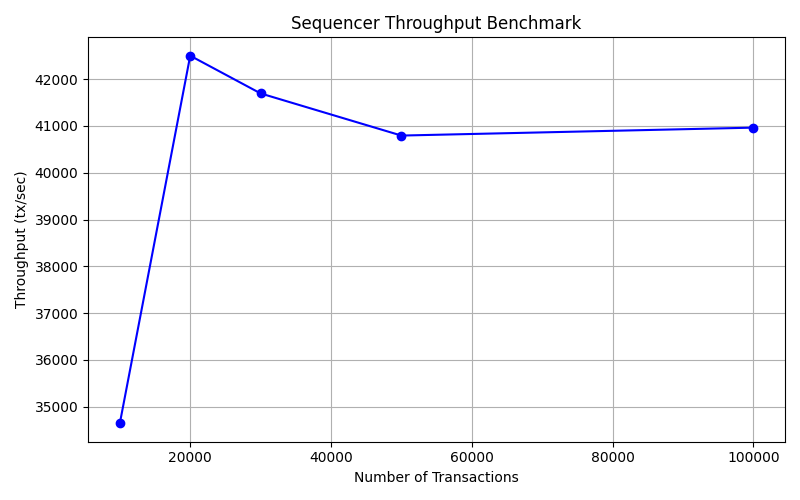}
    \caption{ZK-Rollup Transaction Throughput}
    \label{fig:transaction_throughput}
\end{figure}

To ensure accuracy in the results obtained, the random transaction generation occurs based on seeded randomness such that the generated transactions and their order are the same across multiple runs of the evaluation. This eliminates the possibility of affecting transaction throughput by submitting simpler, more optimized transactions during a run, and complex ones during another. The results in Figure \ref{fig:transaction_throughput} show that the transaction throughput sharply increases to ~42,500 transactions per second (tx/sec) initially and then plateaus to roughly 41,000 tx/sec. This indicates that the sequencer was easily able to handle the lower number of transactions but started to be overloaded as the number of transactions reached 100,000. While the throughput did slightly decrease with a higher number of transactions, it is more than enough to accommodate real-world workloads.

\subsection{Decentralized CRS Generation}

The decentralized CRS generation ceremony was evaluated using a testbed designed to measure both its performance and its fault tolerance. Extensive tests were conducted under normal and adversarial conditions to ensure the robustness of the system. The testbed employed two network configurations, one consisting of 6 nodes while the other consists of 12 nodes, with the CRS being generated at three distinct power sizes (8, 10, and 12). For each combination of node count and power size, 10 independent ceremony iterations were executed to accurately obtain an average for the duration of the ceremony while accounting for network variables.

Furthermore, the resilience of the system against adversarial behavior was accounted for in the testbed by implementing a probabilistic attack model. For each launched configuration, a random number of nodes were chosen to act maliciously. Each malicious node would then decide whether to act maliciously or not during its turn to contribute. If malicious behavior is decided by the node, it can choose one of the following two attack vectors:

\begin{enumerate}
    \item Random disconnection: The node simply disconnects during its turn to simulate a network failure or the node intentionally exhibiting timeouts as a denial of service attack.
    \item Invalid contribution: The node adds an invalid contribution to the CRS during its turn. This can be done by either replacing the CRS completely to disregard previous contributions or by corrupting the contribution.
\end{enumerate}

\begin{figure}[htbp]
    \centering
    \includegraphics[width=\linewidth]{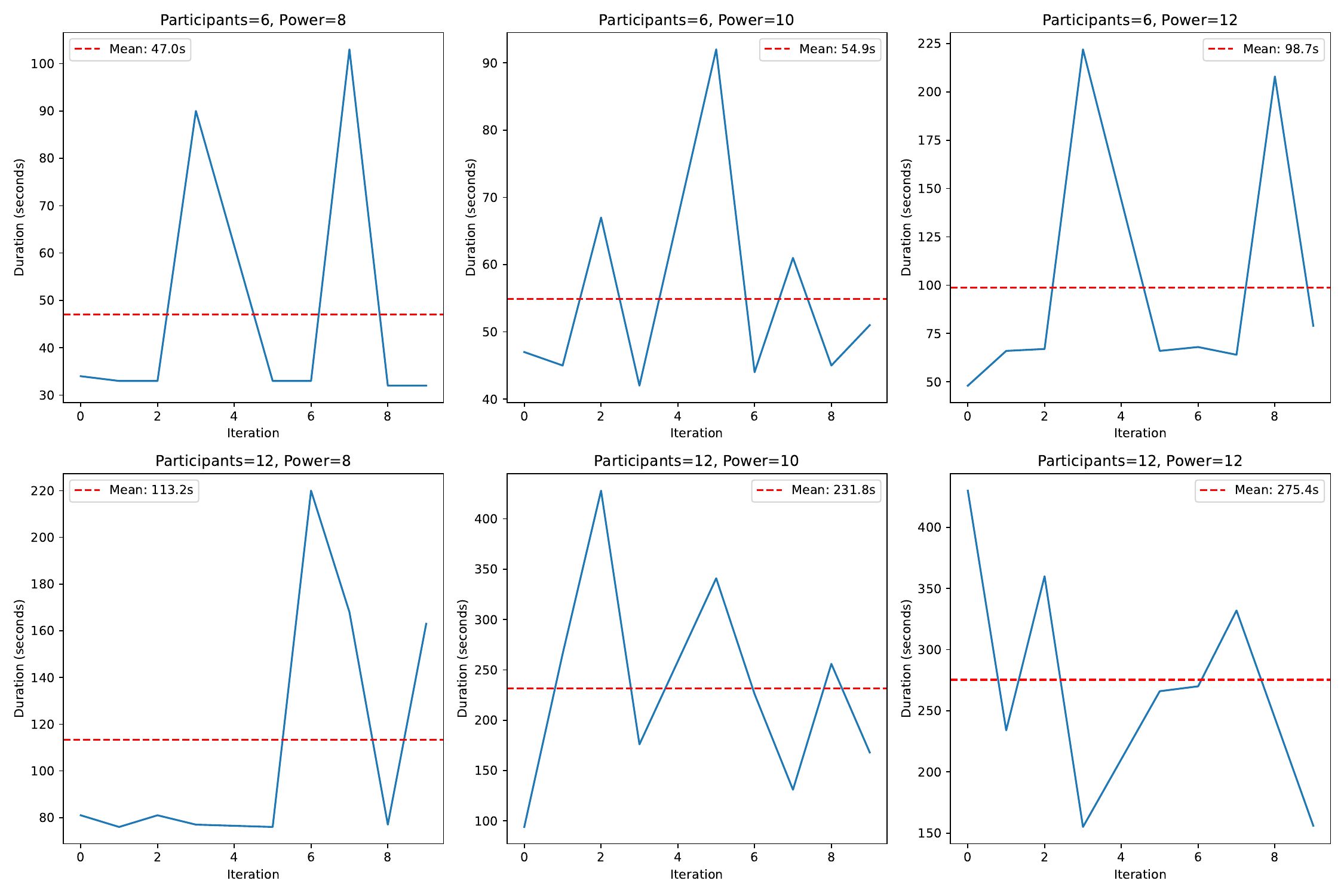}
    \caption{Decentralized CRS Generation Ceremony Durations}
    \label{fig:ceremony_duration}
\end{figure}

The evaluation results are shown in Figure~\ref{fig:ceremony_duration}. The results show that the mean duration of ceremonies increases with increasing power size, rising from 47.0~seconds to 98.7~seconds for the 6-participant configuration and from 113.2~seconds to 275.4~seconds for the 12-participant configuration. Increasing the number of participants also naturally increases the ceremony duration, as additional consensus rounds are required to validate each contribution.

For each participant count and power size combination, no clear trend is observed across individual iterations because each ceremony is executed independently and experiences different adversarial conditions determined by the probabilistic attack model. The variation in ceremony duration is primarily caused by the recovery mechanisms triggered during these attacks. When a participant disconnects or fails to contribute before a predefined timeout, the protocol waits for the timeout period before excluding the participant from the current ceremony and allowing the remaining participants to proceed. Similarly, invalid CRS contributions are detected during protocol verification, rejected through PBFT consensus, and the offending participant is removed from the active participant set. These recovery procedures introduce additional coordination overhead, resulting in longer completion times for some iterations. Nevertheless, all ceremonies complete successfully without requiring a complete restart, demonstrating the robustness of the proposed coordination framework against both transient failures and malicious behavior.

\subsection{Generated CRS Size}

\begin{figure}[htbp]
    \centering
    \includegraphics[width=\linewidth]{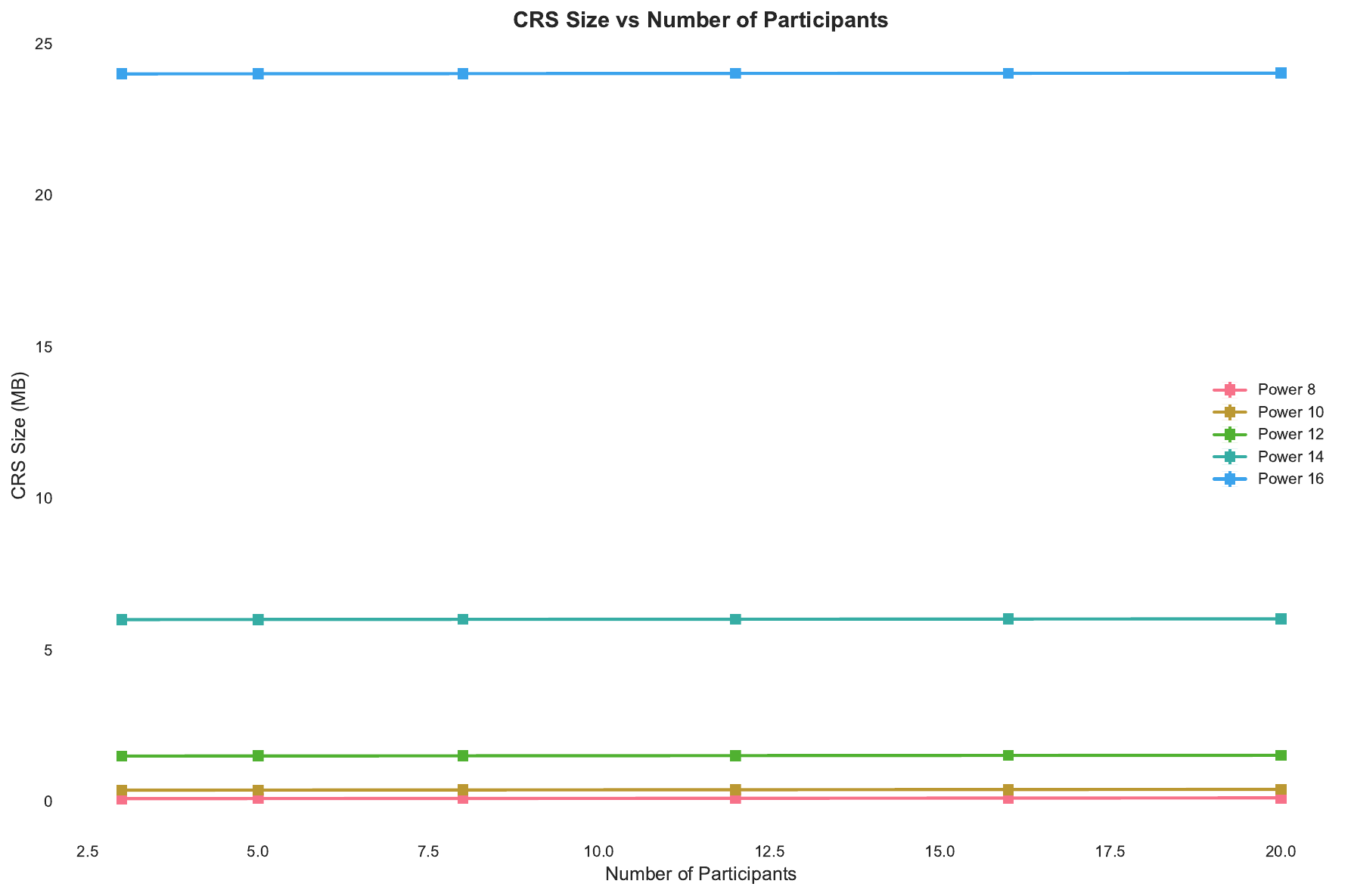}
    \caption{CRS Size Correlation with Number of Participants}
    \label{fig:crs_size}
\end{figure}

We also evaluate the relationship between CRS size, power size, and participant count. Figure \ref{fig:crs_size} shows that CRS size is dominated by the power parameter, with participant count contributing only marginal growth. Consequently, larger participant sets can enhance ceremony security without significantly increasing storage requirements, although they may incur additional coordination overhead.

\section{Conclusion}\label{conclusion}

This paper presented a Layer~2 coordination framework for continuous CRS generation and decentralized CRS lifecycle management. By leveraging a decentralized PBFT-based sequencer network, the proposed framework transforms trusted setup from a one-time cryptographic ceremony into a continuously managed decentralized protocol service while preserving the trust assumptions of multi-party CRS generation. To support diverse deployment environments, the framework provides both smart contract-based and peer-to-peer coordination mechanisms for recurring CRS generation.

Experimental results demonstrate that the decentralized rollup maintains high transaction throughput while enabling recurring CRS ceremonies to complete reliably within practical time bounds, even under adversarial conditions. The proposed recovery mechanisms ensure that transient failures and malicious participant behavior do not prevent successful CRS generation, demonstrating the robustness and practicality of the approach for scalable, trust-minimized zero-knowledge systems.

More broadly, this work demonstrates that decentralized Layer~2 infrastructures can support long-lived cryptographic services in addition to transaction execution, expanding the role of rollups from scalability mechanisms to general-purpose decentralized coordination platforms for blockchain protocols.

\bibliographystyle{IEEEtran}
\bibliography{IEEEabrv, references}

\end{document}